\documentstyle[10pt]{article}
\begin{document}
\begin{center}
{\Large 9.84 {\it vs.} 9.84:}\\ 
{\large The Battle of Bruny and Bailey} \\
\vspace{3mm}
J. R. Mureika\footnote{newt@palmtree.physics.utoronto.ca} \\
{\it Department of Physics,
University of Toronto,
Toronto, Ontario  M5S 1A7~~Canada} \\
\end{center}       

\vskip .25 cm
\noindent
{\footnotesize
{\bf Abstract} \\                        
At the recent 1999 World Athletics Championships in Sevilla, Spain, 
Canada's Bruny Surin matched Donovan Bailey's National and former World
Record 100 m mark of 9.84 s.  The unofficial times for each, as read from the
photo-finish, were 9.833 s and 9.835 s respectively.  Who, then, is the
fastest Canadian of all time?  A possible solution is offered, accounting
for drag effects resulting from ambient tail-winds and altitude.} \\

From the moment Bruny Surin's silver-clad Seville performance popped up
on the screen, it was only a matter of time before the question was raised.\\

This story, of course, has its roots in the July 27, 1996 performance of one
Donovan Bailey, Canada's prime hopeful at the centennial Olympic Games in 
Atlanta.
Coming from behind, Donovan shifted to a gear which that day only he
possessed, surging ahead to the tape in a World Record mark of 9.84s.  Fast
forward to August 22, 1999: Montreal's Bruny Surin battles for global
century supremacy
with the formidable Maurice Greene, who a shy 2 months earlier had eclipsed Bailey's
world mark in an ominously reminiscent 9.79s.  Although falling mildly short of
his rival's 9.80s gold medal romp, Surin's time was far from disappointing,
yet in a way uniquely Canadian: 9.84s.\\

The score so far: 9.84s Bailey, 9.84s Surin-- but is a 9.84s always a 9.84s?  
Without delving too deeply into the philosophical, one needs to address
the standards of electronic timing.  For those not familiar with the 
equipment, the top-of-the-line photo-finish cameras actually sample at 0.001s, 
implying that the athletes' performances
are initially recorded to three decimal places.  For various reasons, the
precision is only kept to two places,
but the rounding process isn't quite scientific.  Unless the
third decimal place is '0', the times are rounded {\it up} to the next highest
hundredth.  So, two performances can be up to 0.009s apart, and {\it still}
be regarded as ``equal''!\\

Herein lies our current dilemma.  In a recent issue of {\it Athletics}
\cite{athletics},
it was pointed out that Bailey's 9.84 s was initially a 9.835 s, while Surin's
9.84 s was really 9.833 s. How does reaction time fit in?  Donovan slept in the 
blocks for 0.174 s (a potential nail-in-the-coffin for an Olympic final!), 
while Bruny blasted ahead of his field in 0.127 s.  So, after a little math, 
Surin clocks in with 9.706 s, but Bailey now leads at 9.661 s.  \\

Is Donovan's performance truly of Olympic proportions, as compared to 
Bruny's??  In case you didn't see this coming (by now you should all know
better), we can't disregard two vital 
pieces of data: wind speed and altitude!  The measurements in question were
$+0.7$ m/s in Atlanta (approximately 315 m above sea level), and
$+0.2$ m/s in Seville (about 12 m above sea level).
Since a tail-wind boosts a sprint time, Bailey's $+0.7$ m/s tail gave him more
of an advantage than Surin's $+0.2$ m/s.  However, a higher altitude sprint is 
easier
than one closer to sea level-- hence, a Seville race will be slower than one in
Atlanta!  What to do??  Is this debate destined for the files of {\it Unsolved
Mysteries}? \\

Through the miracle of numerical modeling, it's possible to estimate the 
benefit associated with each statistic.  Drag is calculated as \\

\begin{equation}
drag = \frac{1}{2} \rho  \; C_d\;  A\;  ( v - w )^2~,
\label{drag}
\end{equation}
where $A$ is the cross-sectional area of the sprinter, $C_d$ is the drag 
coefficient, $\rho$ the density of the air, $v$ the sprinter's speed, and $w$
the wind speed.  Note the dependence on $\rho$ and $w$: the higher the altitude, 
the thinner the air, the lower the value of $\rho$.  Likewise, the stronger the tail 
wind, the smaller
$(v - w)^2$ gets. Hence, both imply a lower overall drag on the sprinter.  
Since the effect of wind will vary with altitude, it's 
reasonable to convert all 
performances to their sea level equivalent (or 0 metres altitude).  
The following chart gives a quick indication of the degree to which a 9.72 s 
sea-level clocking (assuming reaction is 
subtracted) will be boosted by differing wind and altitude conditions.  
The last row represents the elevation of Mexico City, to give appreciation 
for the advantage experience in the 1968 Olympics (the density of air is 
roughly $76-78\%$ that at sea level, so clearly with the right tail wind, 
it's no 
wonder that the sprints and jumps experienced record-breaking performances).\\

\begin{table}
{\begin{tabular}{|r l l l|}\hline
Altitude (m) & $w$: $+0.0$ m/s & $+1.0$ m/s & $+2.0$ m/s  \\ \hline\hline
 0 & $+0.000$ s &$+0.064$ s & $+0.121$ s \\
 1000 & $+0.043$  &$+0.099$  & $+0.149$  \\
 2000 & $+0.079$  &$+0.130$  & $+0.174$  \\
 2234 & $+0.087$  &$+0.137$  & $+0.180$  \\ \hline
\end{tabular}}
\label{correction1}
\end{table}                                                

Plugging the numbers into my model \cite{mymodel}, I find the following
quick figures: the altitude+wind combo for Bailey implies that his race would 
be equivalent to roughly a 9.719 s ($+0.044$ s from just wind; $+0.058$ s 
combined).  Surin's race would correspond to a 9.720 s century ($+0.013$ s
wind; $+0.014$ s combined).\\

There we have it: instead of 9.84 {\it vs.} 9.84, after correcting for
reaction time and drag effects, we end up with 9.720 s {\it vs.} 9.719 s.  
Since exact values of $C_d$ and $A$ are unknown, their estimation introduces a 
degree of uncertainty to any calculation.  So, it's certainly 
not unreasonable to expect that this could account for a 0.001s discrepancy,
implying that Bailey's and Surin's times are effectively indistinguishable! \\

Thus, whose 9.84 s is faster?  According to these preliminary 
results: they're {\it both}
the fastest.  But, with 
two 9.84 s clockings toping the national list, Canada certainly
comes out ahead.  Perhaps 
the 2000 season will shed some definitive light on the individual
battle. \\


\begin{thebibliography}{99}
\bibitem{athletics} Cecil Smith, ``Inside Track'', 
{\it Athletics: Canada's National Track and Field Running Magazine} 
(November 1999)
\bibitem{mymodel} J.\ R.\ Mureika, ``A realistic mathematical sprint model 
accounting for wind and altitude effects'' ({\it in preparation})
\end{thebibliography}
\end{document}